\documentclass[preprint]{aastex}

\usepackage[usenames]{color}                         

\shorttitle{Topology and Observed Features}
\shortauthors{Savcheva et al.}

\begin{document}

\title{The Relation between Solar Eruption Topologies and Observed Flare Features I: Flare Ribbons}

\email{asavcheva@cfa.harvard.edu}
\author{A. Savcheva\altaffilmark{1}, E. Pariat\altaffilmark{2}, S. McKillop\altaffilmark{1}, P. McCauley\altaffilmark{1}, E. Hanson\altaffilmark{3}, Y. Su
\altaffilmark{4,1}, E. Werner\altaffilmark{5} \& E. E. DeLuca\altaffilmark{1}}
\affil{\altaffilmark{1}Harvard-Smithsonian Center for Astrophysics, 60 Garden Street,
Cambridge, MA 02138, USA; \\ 
\altaffilmark{2}LESIA, Observatoire de Paris, CNRS, UPMC, Universit\'e Paris Diderot, 92190 Meudon, France\\
\altaffilmark{3}Department of Physics, University of California, Berkeley, CA 94720\\
\altaffilmark{4}Key Laboratory for Dark Matter and Space Science, Purple Mountain Observatory, Chinese Academy of Sciences, Nanjing 210008, China\\
\altaffilmark{5}Department of Physics and Astronomy, Uppsala University, Uppsala, SE-751 20 Uppsala,  Sweden}
\email{asavcheva@cfa.harvard.edu}

\begin{abstract}
In this paper we present a topological magnetic field investigation of seven two-ribbon flares in sigmoidal active regions observed with {\it Hinode}, {\it STEREO}, and {\it SDO}. We first derive the 3D coronal magnetic field structure of all regions using marginally unstable 3D coronal magnetic field models created with the flux rope insertion method. The unstable models have been shown to be a good model of the flaring magnetic field configurations. Regions are selected based on their pre-flare configurations along with the appearance and observational coverage of flare ribbons, and the model is constrained using pre-flare features observed in extreme ultraviolet and X-ray passbands.
We perform a topology analysis of the models by computing the squashing factor, $Q$ in order to determine the locations of prominent quasi-separatrix layers (QSLs). QSLs from these maps are compared to flare ribbons at their full extents. We show that in all cases the straight segments of the two J-shaped ribbons are matched very well by the flux-rope-related QSLs, and the matches to the hooked segments are less consistent but still good for most cases. In addition, we show that these QSLs overlay ridges in the electric current density maps. This study is the largest sample of regions with QSLs derived from 3D coronal magnetic field models, and it shows that the magnetofrictional modeling technique that we employ gives a very good representation of flaring regions, with the power to predict flare ribbon locations in the event of a flare following the time of the model.

\end{abstract}

\keywords{Sun:magnetic fields --- Sun: sigmoid --- Sun: flares ---
Sun: X-rays}
 
\section{Introduction} \label{sec:Intro}

Solar flares, often accompanied by coronal mass ejections (CMEs), are the most powerful phenomena originating from the variable sun. The energy powering these events is thought to come from magnetic reconnection, a process that converts free magnetic energy into radiation, energetic particle acceleration, and kinetic energy of plasma \citep{Forbes06}. Various observed flare and CME-related features can be understood in the context of the standard 2D flare model, the basics which were laid out by \cite{Carmichael64}, \cite{Sturrock66}, \cite{Hirayama74}, and \cite{Kopp76}, and further extended based on {\it Yohkoh} observations by \cite{Shibata95} and \cite{Tsuneta96,Tsuneta97}. In this model, flare ribbons correspond to  the chromospheric footpoints of newly-reconnected field lines, and flare loops represent reconnected field lines cooling beneath the reconnection site at the base of the escaping flux rope. While the 2D model has successfully explained multiple related phenomena, flares are intrinsically 3D events. The physical processes underlying flares, particularly magnetic reconnection, may have properties in 3D that significantly deviate from the 2D view \citep{Aulanier06b,Janvier13}. For example, the shape of the ribbons and the existence of transient coronal holes (dimmings), associated with the footprints of the escaping flux rope, are only explained by the 3D models.

It is therefore necessary to better understand the 3D properties of ultraviolet (UV) flare ribbons because of their complex morphology and dynamics offer a particularly straightforward way to study the three-dimensionality of flares. UV flare ribbons are produced by heat conduction and energetic particles streaming down the flare loops, which interact with the denser layers of the solar atmosphere \citep{Fletcher11}. Flare ribbons arise in the process of chromospheric evaporation, whereby chromospheric heating from the corona is suddenly increased and overcomes the cooling rate of the chromospheric material, dramatically increasing the density and temperature of upflowing plasma \citep{Antonucci84}. This process can be aided by particle precipitation from the reconnection site, as discussed in \cite{Reid12}. Because heat conduction fronts and particles flow down magnetic field lines, the shape, separation, and size of flare ribbons are intimately related to the magnetic structure under the flare site \citep{Yurchyshyn00}. 

Although chromospheric evaporation is included in the 2D standard model, both for producing ribbons and populating the flare loops with dense plasma, a 2D view cannot facilitate predictions about the shape of flare ribbons. When extending the standard model concepts to 3D, the shape of the ribbons are supposed to be determined by the footpoint locations of reconnected field lines, along which energy is flowing down. The shape and dynamics of the flare ribbons are thus tightly linked with the 3D magnetic configuration and dynamics during a flare. Ribbons can therefore be used to identify properties of the reconnection site, to gain insight into the particular flare configuration, and to test 3D flare models \citep{Gorbachev88}.

Magnetic topology aims to studying and identify specific locations in the 3D magnetic field where the strong electric current sheets necessary to induce reconnection are likely to form. This enables us to determine the field lines that will be involved in reconnection and to predict along which field lines energy will flow to form flare ribbons. A separatrix surface, through which the magnetic field connectivity is discontinuous \citep{Longcope05}, is such an example of a magnetic configuration preferential for magnetic reconnection. The first to match ribbons to separatrices from a potential field model of an active region were \cite{Gorbachev88}. Overlap between flare ribbons and topological features in the CME domain has previously been demonstrated for true separatrices on the photosphere from potential and linear force-free field (LFFF) models \citep{Mandrini91, Mandrini95, Mandrini14, Demoulin93, Demoulin94, vanDriel94}. A null point is another example of a topological feature exhibiting a distinctive shape: observations showing a circular ribbon and remote brightening  reveal the existence of a 3D null point magnetic configuration with a fan-spine topology \citep[e.g.][]{Masson09, Masson14}.

Apart from atypical cases \citep[][]{Liu14,Dalmasse15}, eruptive flares tend to involve the formation of two flare ribbons, which generally form two `J' shapes \citep{Schmieder97, Chandra09, Schrijver11} that are oppositely directed, with the straight part of the Js adjacent to each other on either side of the polarity inversion line (PIL) and the hooks facing each other on the opposite ends. Understanding the magnetic topology associated with these two J-shaped ribbons can enlighten us on the processes leading to and developing during eruptions. A subcategory of the two J-shaped ribbon flares are those for which the active region of origin presents a forward or inverse S-shape in its extreme ultraviolet (EUV) or X-ray emission for some time prior to the flare. These regions are commonly known as sigmoids \citep{Rust96}. Sigmoid-region two J-shaped ribbon flares have been interpreted to occur in bi-polar configurations containing a flux rope. Evidence for the presence of flux ropes comes from magnetic modeling as well as the very shape of the ribbons. Non-linear force-free field (NLFFF) models and extrapolations have been used to reveal the presence of a twisted flux rope \citep{Schrijver06, Savcheva09, Savcheva12a, Savcheva12c, Liu14}. Additionaly, the J-shape of the ribbons has been shown to correspond to the specific topological structures associated with a twisted flux rope: quasi-separatrix layers (QSLs) and hyperbolic flux tubes (HFT). 

QSLs are purely 3D topological structures that do not exist in 2D. They correspond to subvolumes of the magnetic field where the connectivity of neighbouring field lines, while continuous, experiences drastic changes \citep{Priest95,Demoulin96b, Titov02, Titov07,Pariat12}. QSLs, as separatrices, are preferential sites for the development of magnetic reconnection \citep{Demoulin96b,Aulanier05b,Aulanier06b,Janvier13}. HFTs are located at the intersection of QSLs  
and are a generalization of magnetic separators \citep{Titov02} that are typical of the regions below twisted flux ropes \citep{Titov02,Savcheva12a,Savcheva12b,Pariat12,Janvier13,Guo13,Zhao14}. Interestingly, the photospheric footprints of the QSLs surrounding the twisted flux rope present the shape of two inverted Js \citep{Titov02,Savcheva12b,Pariat12}. Magnetic reconnection occurring at these QSLs would thus naturally lead to the generation of the two J-shaped ribbons observed during flares. Observations of two J-shaped flare ribbons are therefore strong evidence for the presence of a twisted flux rope \citep{Titov02,Schrijver11,Savcheva12b,Zhao14, Dudik14}. 
J-shaped ribbons have also been associated with observations of J-shaped electric currents during flares, which is theoretically predicted by the 3D standard model since the QSLs are preferential sites for current accumulation \citep{Janvier14}. Finally, reconnection developing at 3D QSLs, slipping reconnection \citep{Aulanier06b}, differs from 2D cut-and-paste reconnection. Motions of UV brightening observed during the formation of the J-shaped ribbons have exhibited slipping behaviors \citep{Dudik14}, further confirming the presence of J-shaped QSLs in sigmoid flare events. 

However, most previous studies have only examined case-by-case connections between J-shaped ribbons and magnetic topologies, and no statistical studies have been done. A selection bias potentially exists for regions chosen based on their favorable topological properties, i.e. only regions that do present a close correlation between J-shaped ribbons and J-shaped QSLs have been studied and published on. In this study, we find particular topologies for a sample of well-observed events. We present a correspondence between the morphology of the photospheric QSLs computed from observationally-constrained magnetic field models of the coronal field and the flare ribbons for 7 sigmoid region eruptions. These regions have been selected based on their X-ray and EUV morphology prior to the flare and on the shape of the ribbons. No topology criteria have been used, and we therefore reach a more statistically significant conclusion on whether or not J-shaped flare ribbons in sigmoid regions tend to match the distribution of QSLs. This study allows us to control whether a portion of J-shaped ribbons is better matched to the coronal magnetic field model than others. Such comparisons with observed ribbons enables us to study and control the quality of our magnetic field model in a new and innovative way.

This study is the observationally-constrained counterpart of the simulation results of  \cite{Aulanier12} and \cite{Janvier13}, who use QSLs to extend the standard flare model to 3D and the next step after \cite[][; hereafter S12]{Savcheva12b}. The ultimate test for whether our approach is useful or not is the degree to which the flare ribbons match the QSLs derived from unstable models for erupting sigmoidal active regions. The validity of this approach is discussed in Section\,\ref{sec:NLFFF}. The paper is organized as follows: In \S2, we present the observations of the erupting sigmoidal active regions. In \S3, we discuss the specifics of the magnetic modeling, and \S4 details the QSL calculation. In \S5, we present our main results about the association of QSLs with flare ribbons. Concluding remarks, along with a preview of Paper II, are given in \S6. 

\section{Observation} \label{sec:Obs}

In \cite{Savcheva14}, we presented a sigmoid catalog spanning the duration of the Hinode and SDO missions\footnote{Hinode \& SDO Sigmoid Catalog: \url{http://aia.cfa.harvard.edu/sigmoid/}}. For this paper, we have chosen seven sigmoidal active regions selected from the catalog. We use the recently-proposed naming convention for solar targets endorsed by the {\it Astrophysical Journal, Solar Physics}, and the International Astronomical Union, under which our 7 regions are labeled:\\ 
SOL2007-02-12T07:30:00L075C297\\ 
SOL2007-12-07T04:20:00L085C296\\
SOL2010-04-08T02:35:00L110C176\\ 
SOL2010-08-07T17:55:00L100C002\\ 
SOL2012-05-08T09:21:00L108C232\\ 
SOL2013-03-15T05:46:00L116C077\\
SOL2013-04-11T06:55:00L110C071\\

Two regions, SOL2007-02-12 and SOL2007-12-07, were thoroughly observed by the X-ray Telescope on Hinode \citep[XRT;][]{Golub07} and STEREO/EUVI \citep{Wuelser04}, while the rest were observed during the Solar Dynamics Observatory (SDO) mission and have detailed observations from the Atmospheric Imaging Assembly \citep[AIA;][]{Lemen12}. 

Regions were chosen based on the following criteria: 
\begin{enumerate}
\item Being eruptive sigmoidal regions, which  are taken as a direct evidence for the existence of magnetic flux ropes before the eruptions \citep{Green11}.
\item Having very good observational coverage around the flare time.
\item Having well-defined double ribbons on both sides of the PIL, i.e. having two-ribbon flares. 
\item Having observations of other flare- and CME-related features, such as transient coronal holes (dimmings) and flare arcades.
\end{enumerate} 
This sample does not include all of the catalog regions that fit our criteria. The case-by-case numerical modeling and topology analysis that we performed on each region (see Sections \ref{sec:NLFFF} \& \ref{sec:QSLmeth}) represents a substantial and time consuming effort. While this sample of seven regions remains statistically limited, it far exceeds similar published analyses of flaring regions (cf. Sect.\,\ref{sec:Intro}). 
  
It is important to emphasize that we select regions based purely on the quality of the data  and on the observational properties of the EUV and X-ray emission before and during the flare. None of our selection criteria are directly based on the magnetic field properties of the region. The selection does not presume how well the magnetic models fit the data or how well the ribbons match the QSLs. The topological analysis was performed blindly, using only the magnetic field model and without taking in to account the EUV and X-ray emission of the individual region. We present all regions in the original sample, some of which are not perfectly matched. The list of the regions by date, including the time ranges over which different flare and CME features were observed, are presented in Table\,\ref{tbl-1}. 

As can be seen from Table\,\ref{tbl-1}, our sample contains both weak (B- and C-class) and some stronger flares (up to M6), so we can demonstrate the correspondence of QSLs and ribbons in different types of magnetic environments. Both STEREO flares are B-class flares that do not display clear motion of the ribbons, probably due to the low intensity of the flare. Our sample does not contain any X-class events because all suitable X-class flares have excessive saturation 
that precludes a clear examination of the flare ribbon evolution.

In this study, we are concerned with the initial moment when the full lengths of the flare ribbons appear. This moment is defined by eye by inspecting image sequences in several AIA or STEREO channels. The ribbons appear at different times and to different extents in the various channels since they are sensitive to different temperatures. For the SDO regions, we rely primarily on the 304\,\AA\, channel since it is often less saturated and the ribbon shape is not confused by overlying loops as in the other channels. We did not make the traditional choice to use 1600\,\AA\,\ images because not all flare ribbons were seen in their entirety in this channel and because the ribbons generally appeared later in the 1600 \AA{} channel (i.e. the 304 \AA{} channel captures hotter plasma than the UV channels), and we wanted to obtain a moment as close to the beginning of the flare as possible. Because of the more limited observation cadence (5\,min) for the STEREO regions, we chose the channel that showed these particular ribbons the best, i.e. the 195\,\AA\, channel. In Fig.\,\ref{Ribbons_AIA_1} and \ref{Ribbons_AIA_2}, we display the flare ribbons in all regions at the times when the ribbons are seen first at their full extent. This time, $t_{ribbons}$ (also listed in Table\,\ref{tbl-1}), is the reference time used in Section\,\ref{sec:QSLCME} to make the overlays between the QSLs and flare ribbons. 

Some ribbons brighten more or less simultaneously over roughly their whole length and others grow in length as time progresses. The moment, $t_{ribbons}$, that we use to compare the ribbons with the magnetic model is the first moment when the ribbons are observed over their full extent. In classical two-ribbon flares, ribbons display two kinds of motions \citep{Asai02, Asai04, Fletcher09,Qiu09, Qiu10}. One is a ``zipper'' motion (fast elongation) parallel to the PIL during the impulsive phase of the flare with velocities of 10-100\,km\,s$^{-1}$. The other is an expanding (separating) motion perpendicular to the PIL during the gradual phase of the flare with velocities of 1-20\,km\,s$^{-1}$. At $t_{ribbons}$, most of the zipper motion has taken place, but the expanding (separating) motion has not started or is at a very preliminary stage. The ribbons in both of the STEREO events did not show significant motion over the duration of the flare, probably because of the weaker flare intensity. For these events, because of the lower cadence, it was also harder to strictly pinpoint $t_{ribbons}$. As can be noted from Table\,\ref{tbl-1}, $t_{ribbons}$ is usually within 10\,min after $t_{GOES}$, of the flare start time measured by GOES.

\begin{figure}[h!]
\epsscale{1.}
\plotone{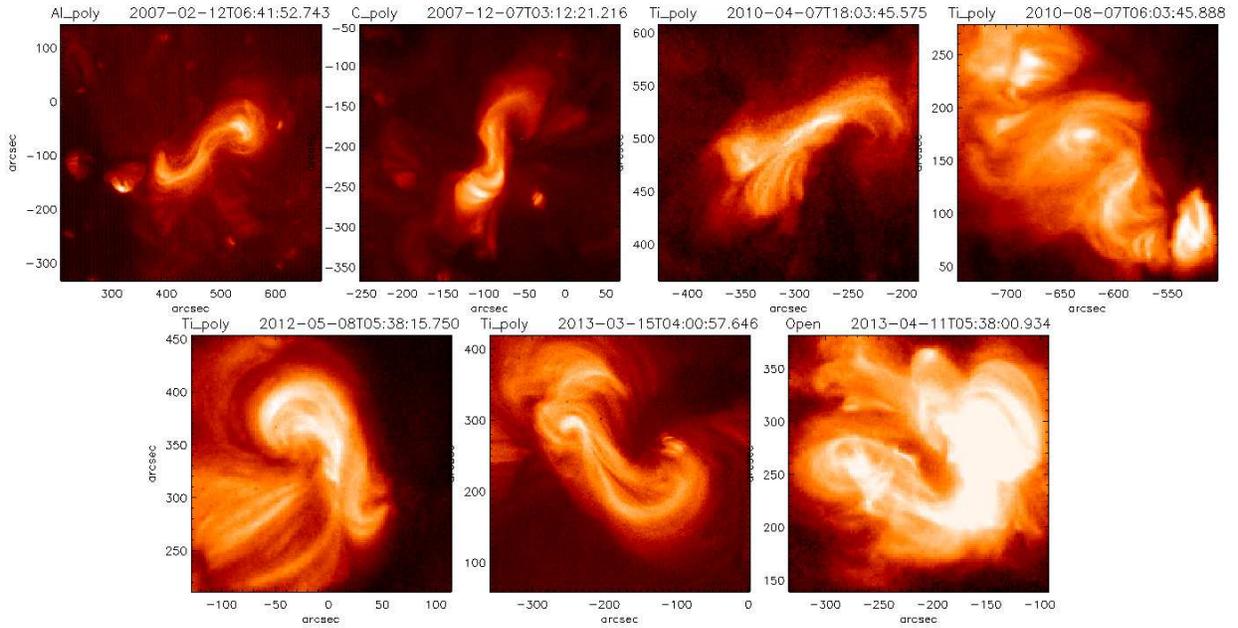}
\caption{X-ray images from XRT used to fit the NLFFF models for all regions. Notice the S-shaped and 2J-shaped loops in most regions that give the sigmoidal shape.    
\label{XRT}} 
\end{figure}

In Fig.\,\ref{XRT}, we show pre-flare observations from XRT for each of the seven regions, revealing their sigmoidal nature. The XRT observations are displayed close to the time, $t_{mag}$ (listed in Table\,\ref{tbl-2}), at which the 3D magnetic field models have been produced for each region. Most images show S-shaped loops that outline the sigmoids. All sigmoids on the figure are long-lasting with the exception of SOL2013-04-11, which was transient and appeared a couple of hours before the GOES flare. For this region, the XRT image is not taken at a time useful for constraining the model, i.e. no S- or J-shaped loops are visible, so we use AIA 335\,\AA{} instead. 

As we will discuss in Section\,\ref{sec:NLFFF}, the models are constrained by a number of coronal loops traced from these images. The images were chosen to show coronal structures that did not show any flare-related changes yet but were also representative of the immediate pre-flare configuration. For example, some of the loops seen in the figure lifted up as the flare was starting and others faded away, giving place to the flare arcade.

\section{Magnetic Field Modeling} \label{sec:NLFFF}

Since these regions are sigmoidal, we can gain insight into their 3D magnetic field structure by building NLFFF models, which are the only models that can represent the core field of a magnetic flux rope and the overlying potential arcade \citep{Savcheva09}. We assume that the coronal field is force free, hence that $\nabla\times\mathbf{B}=\alpha\mathbf{B}$, where $\mathbf{B}$ is the magnetic field and $\alpha$ is the torsional parameter that is constant along a given field line. A force-free magnetic field is ``non-linear'' when the torsion parameter is a function of position. Different field lines have different values of $\alpha$, and $\alpha$ is as close to a constant along the field line as possible. In our models, the cores of the sigmoids are represented by sheared and twisted fields, while the outlying fields have a nearly potential configuration. The relaxation method ensures that the resulting field is force-free and non-linear. 

We have used the flux rope insertion method \citep{vanBallegooijen04} to produce models of the magnetic field in each of our observed cases for the instants, $t_{mag}$, preceding the eruptions by about 0.5-1\,hour (see Table\,\ref{tbl-1} and \ref{tbl-2}). Briefly, the flux rope insertion method consists of the following steps: 1) A global potential field extrapolation is performed based on a low-resolution synoptic SOLIS magnetogram. 2) A region of interest is selected from a high-resolution MDI or HMI magnetogram. 3) A modified potential field extrapolation is performed in the high resolution region with side boundary conditions set by the global potential field extrapolation. 4) A flux rope is inserted following the path of an EUV filament seen STEREO 195\AA\, or AIA 304\AA\, images. 5) A grid of models is created with different combinations of axial and poloidal fluxes. 6) The configurations are relaxed to a force-free state using magnetofrictional relaxation with some amount of hyperdiffusion, which acts to smooth gradients in $\alpha$. 7) Field lines from each model are matched to observed coronal loops selected from X-ray or EUV (94\AA\, or 335\AA\,) images. We choose loops from the core of the active region where the field is strongly non-potential, usually representing the spine and elbows of the sigmoid, as well as any linear, sheared, and more potential loops, thus spanning the whole magnetic configuration as well as possible. Consequently, this method produces a 3D magnetic field structure that is heavily constrained by observations. The details of the method are further discussed in \cite{Su11} and \cite{Savcheva12a}. We have a quantitative measure of the goodness-of-fit for each model. The goodness of fit criteria are explained extensively in \cite{Savcheva09}. By looking at the range of solutions near the best-fit model, we can estimate the robustness of the best fit, or how sensitive the models are to the choice of initial parameters.

The instances for which we build the models correspond to the times just before the first flare-related morphological changes start to appear in the AIA/STEREO images. For all regions, we calculate a grid of magnetic models covering a range of flux rope axial and poloidal fluxes, as we did in \cite{Savcheva09} and \cite{Savcheva12c}. We have chosen the grid to cover models with little flux (e.g. $\Phi_{\mathrm{axi}}=1\times10^{20}$\,Mx and $F_{\mathrm{pol}}=5\times10^9$\,Mx\,cm$^{-1}$) that produce sheared arcades to very highly twisted (large amount of polidal flux, e.g. $F_{\mathrm{pol}}=1\times10^{11}$\,Mx\,cm$^{-1}$) and/or sheared (large amount of axial flux, e.g. $\Phi_{\mathrm{axi}}=5\times10^{21}$\,Mx in some cases) fields that exceed the stability limit for the particular region. \cite{Su11} showed that the stability of the configuration is tightly controlled by the amount of axial flux and that the fit of the models to the images is weakly dependent on the poloidal flux. We issued best-fit models that closely match the observed loop configurations in XRT synoptic images and AIA 335\AA\ and 94\AA\ channels when XRT was not available. This shows that our models are robust enough to reproduce excellent fits to high resolution AIA observations in the core of active regions. The first such NLFFF model constrained by AIA observations was provided by \cite{Su11}.

The flux rope insertion method produces models that are on a spherical wedge grid. However, the QSL computation is performed in Cartesian coordinates, so we transform the spherical to Cartesian coordinates of the grid as we did in S12.

Although in previous studies we were concerned with the best fit models that represented the equilibrium coronal magnetic field preceding flares \citep{Su11,Savcheva09,Savcheva12c}, here (and in a forthcoming paper) we are more interested in understanding how the field evolves from a static to a dynamic flaring state and in what the magnetic field structure is at the flare onset. The flux rope insertion method produces a stable relaxed model when the forces on the flux rope balance. If no balance is found, the solution continues to evolve as the number of iterations increases. \cite{Kliem13} demonstrated that the stability boundary identified in our NLFFF code is reproduced by a full 3D MHD solution. This gives us confidence that the marginally stable solutions found by our NLFFF method are reasonable representations of the magnetic field at the onset of the flare instability.  

As discussed earlier, if too much flux is used in the flux rope, a relaxed equilibrium state may never be reached by the magnetofrictional relaxation \citep{Su11} and the model is unstable. The flux rope in these unstable models continues to expand and rise with continued iterations, as demonstrated in \citet{Su11}. The unstable models are hence not force-free since since residual forces exist in the volume that drive the expansion of the flux rope. In most cases, the best-fit model based on the pre-flare configuration is unstable or marginally stable. In the marginally stable cases, we add a unit of axial flux  (1-2$\times10^{20}$\,Mx) to the axial flux of the best-fit model, which pushes the models over the edge of stability.

\begin{figure}[h!]
\epsscale{1}
\plotone{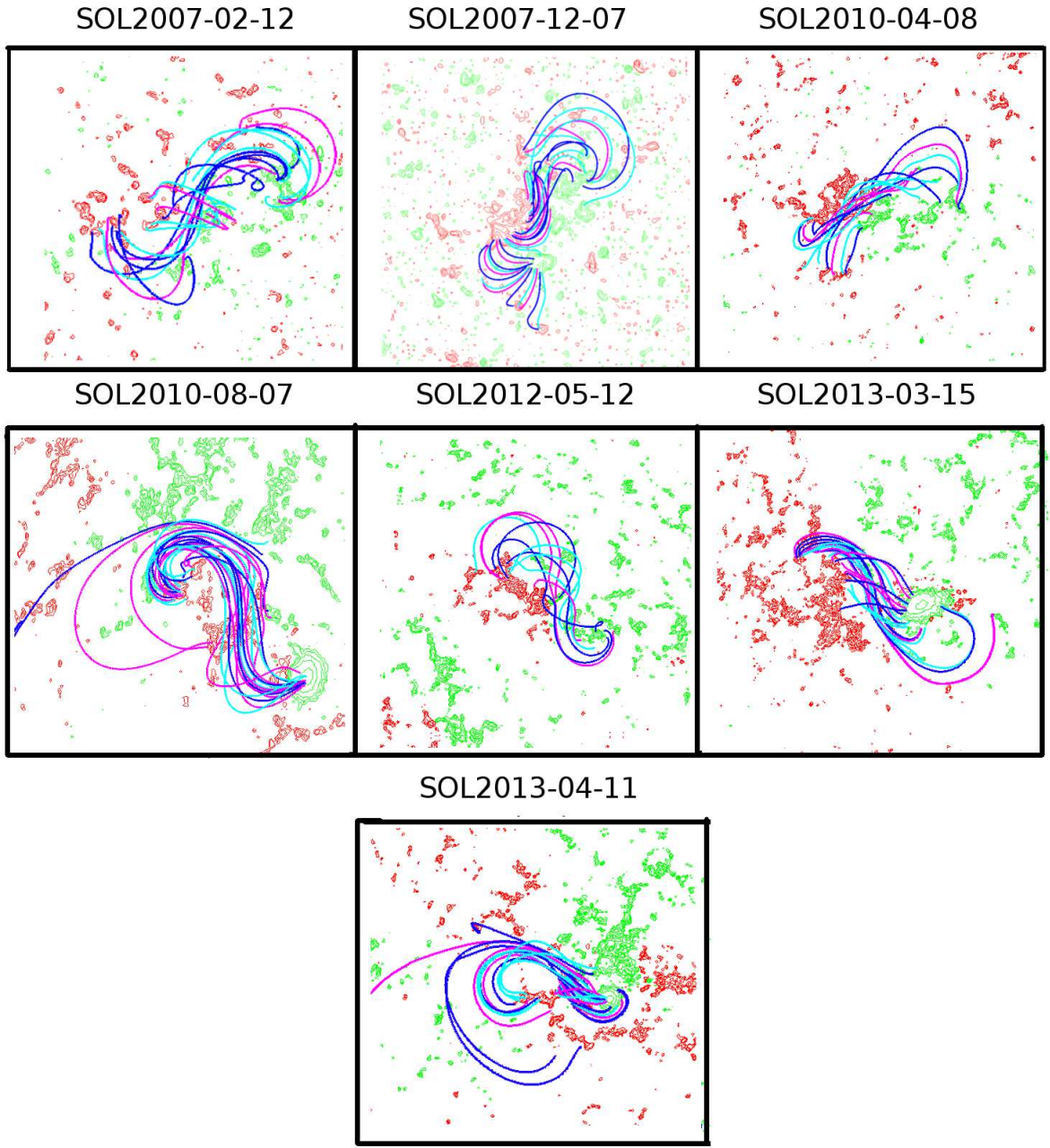}
\caption{Some sample field lines traced from the best-fitting unstable models for all regions studied representing the 3D magnetic field structure. The red and green contours mark the positive and negative flux distributions in the photosphere respectively. The loops are overlaid onto the XRT images used to fit the models (see Fig.\,\ref{XRT}).
\label{FLs}} 
\end{figure}

There are usually no suitable coronal loops that can be used to match the unstable model to the data. The flare loops are not a good choice for fitting because they undergo the strong-to-weak shear transition \citep{Aulanier12}, and the match will depend on the iteration number, which is a central topic of Paper II (see for details). We choose the closest unstable model to the best-fit model and further relax it with no form of diffusion except a numerical one. If the model is stable with the current relaxation scheme we are using, it converges to a NLFFF state at about a relaxation iteration number of 30,000. We save every 10\,000-15\,000 iterations in order to track the evolution of the magnetic configuration with further relaxation. In the case of the unstable models, the flux rope keeps expanding and forms an HFT underneath it \citep{Titov07, Savcheva12a,Savcheva12b}. In Table\,\ref{tbl-2}, we have given the times, input and output parameters of the unstable models we use for all regions. The input parameters are the axial flux ($\Phi_{axi}$) and the poloidal flux per unit length ($F_{pol}$), which when multiplied by the length of the flux rope path results in the total poloidal flux given in the table, $F_{pol,tot}$. The output parameters are potential field energy ($E_{pot}$), computed before the introduction of the flux rope, the free energy ($E_{free}$), computed by subtracting the total energy of the NLFFF field and $E_{pot}$. The last column is the relative helicity, computed from equation (B2) from \cite{Bobra08}. Sample field lines from the best fitting models are shown in Fig.\,\ref{FLs}. One can compare Fig.\,\ref{XRT} and Fig.\,\ref{FLs} to see that there is a very good qualitative match of the field lines to the X-ray emission. 

In principle, modeling an unstable eruptive configuration with NLFFF is not easy, even when using vector fields to extrapolate the coronal field. This is in part because the photospheric magnetic field changes only very subtly in response to the flare, and this change is mostly observed in the transverse component of the vector field \citep{Sun12,Wang12}. In our case, we use LoS magnetograms in which the change is virtually undetectable. During the relaxation process, the vertical field distribution $B_z$ at the photosphere stays fixed and equal to the pre-eruptive LoS magnetogram that we used to obtain the best fit model. The $B_x$ and $B_y$ at the photosphere stay equal to zero in the boundary layer during the relaxation, enforcing strict line tied conditions and preserving the initial purely vertical boundary condition. Even if the change in the LoS field during a flare was apparent, it would be most probable that we cannot choose the initial model and relaxation parameters precisely enough to represent this change in the final 3D field. There have been studies \citep{Jiang12,Liu14,Zhao14} of the quasi-static evolution of the magnetic field covering the time of a flare by utilizing vector magnetogram extrapolations, which in principle can capture the flare-related changes in the configuration. In \cite{Zhao14} they show the existence of a flux rope before and its absence after the flare. There is little discussion of how the flare-related changes to the photospheric vector field affect the coronal field resulting from the extrapolation. This is a fundamental difference between the flux rope insertion method and the vector NLFFF extrapolations in that after having identified a stable or marginally stable solution that matches the observations, we can perturb that solution so it becomes unstable.
  
Although the unstable models are not fully self-consistent, using a marginally unstable model allows for a description of the flaring magnetic configurations close to the expected real 3D magnetic field in each regions at $t_{ribbons}$. \cite{Kliem13} performed a data-constrained MHD simulation initiated by the \cite{Su11} NLFFF models as initial conditions and showed that the dynamics of unstable models in the magnetofrictional relaxation is close to the dynamics of an unstable flux rope in the MHD simulation. The magnetofrictional relaxation, being far more economic in numerical resources than an MHD simulation, allowed us to study marginally unstable models for all seven studied regions, which would have been computationally prohibitive using an MHD simulation. The marginally unstable models therefore allow us to obtain an economic and satisfactory representation of the {\it flaring} active region field.

It is important to emphasize that although possible, we do not use any additional free parameters or constraints on the models based on the eruptive or post-eruptive coronal or magnetogram structures. All parameters and constraints are applied using the pre-flare configuration. We are using the predictive capabilities of the standard model to test the quality of our reconstruction method. The topological analysis of our models, which was performed independently of the model production, allows us to predict the location of the ribbons. If the ribbons do not match the QSLs at all, we cannot say anything about the eruptive structure of the region and the reconnection site. On the other hand, if there is a good qualitative match, we can be confident that we have found a solution that represents the observations well and supports the validity of our theoretical prediction.

\section{The QSL calculation}\label{sec:QSLmeth}

As mentioned in the introduction, based on previous work \citep[e.g.][]{Savcheva12b,Liu14,Janvier13,Janvier14} and the standard flare model, we expect that the flare ribbons match the QSLs in the low corona. In order to determine the fits between the magnetic model and the observed ribbons, we perform a topological analysis of the 3D magnetic field data by computing the squashing factor ($Q$) \citep{Titov02} in different 2D cuts of the domain. To compute these maps, we use the method of \cite{Pariat12} that was successfully applied to a single NLFFF model in S12. This is an iterative method that can take any cut through the domain and first compute a map of the QSLs with the original resolution of the model, which is $1.5\times10^{-3}R_{\odot}$, followed by increasing the resolution where $Q$ is high and repeating the process until convergence to a similar value of $Q$ is reached. 

For each unstable model of the studied regions we, calculate the Q-maps at four heights in the corona: 2100\,km, 3150\,km, 4200\,km, and 5250\,km ($z=2-5$ in simulation cells). The reference boundary is at $z=2$, since below this height the intrinsic complexity of the magnetic field is too large, i.e. a large number of bald patches that exist make the converging QSL computation very slow, as discussed in S12. Taking the reference boundary for the QSL computation as the photospheric layer only focuses on topological features that are situated very low and prevents clear localization of the topological structures that extend into the corona and are involved with the large-scale flare dynamics. Taking the reference field a few pixels above would correspond to a chromospheric-like region where the plasma beta would be closer to unity but where line tying would still be effective. This also allows us to pick out significant topological features in the quiet sun where the field is usually very week, highly fragmented, and containg interspersed polarities. 

Once the $Q$ factor has been computed from an unstable model based on magnetic field data obtained at $t_{mag}$, we overlay each of these QSL maps with an image of the flare ribbons at $t_{ribbon}$. The overlay is performed in such a way that the Cartesian QSL map is projected onto a sphere and the height at which the QSL map is calculated is taken into account in the projection. We overlay each map for the four different heights onto the image by shifting the QSL map coordinates forward in time to match the time difference between the moment when the model was computed and the time of the flare ribbon image, which in some cases was up to two hours. In addition, we use prominent magnetic features, such as sunspots and large, peculiarly-shaped magnetic concentrations to further adjust the alignment. The co-alignment precision can vary between different location in each co-aligned image because of the transformation from the spherical data to the Cartesian one. Hence, for each studied region, the coalignment may be better in a portion of the image than another. Overall, we estimate that this process gives us an alignment precision of about 5-10 AIA pixels or 3$^{\prime\prime}$-6$^{\prime\prime}$ on average. 

Given the complexity of both the QSL maps and the flare ribbon observations, the existence of a large number of QSLs related to small scale features, and the presence of multiple secondary brightenings, we only focus on a qualitative comparison of the QSL maps with the ribbon morphology. We determine the quality of the fit by eye employing a few basic criteria. The QSLs should first capture the portion of the ribbons parallel to the PIL (the straight part of the J-shaped ribbon). This section is indeed directly related to the extent and height of the HFT located below the erupting flux rope. Its dynamics are related to the rise of the flux rope and the reconnection developing along in the CME current sheet. Good alignment between the QSLs and the flare ribbons indicates that the central axis of the twisted flux rope is located at the right height and follows the right shape. Flare ribbons are frequently observed to not be fully facing each other, with positions slightly shifted in the direction of the PIL. This shift is related to the shear present in the pre-eruptive configuration. If the QSLs are able to capture such shift in position, it indicates that the modeled field contains the right amount of shear around the flux rope. The second criterion to evaluate the quality of the fit is the hook of the J-shaped flare ribbons, the sections going away from the PIL. The position of the hook is related to the position of the central axis of the twisted flux rope \citep{Pariat12,Savcheva12b,Janvier13}. Its shape is related to the amount of twist present in the rope: a more twisted flux rope will present a more pronounced hook, eventually closing in on itself for very large twist. Finally, we hope to capture smaller scale detail in the shapes of the ribbons. 

\begin{figure}[h!]
\epsscale{0.8}
\plotone{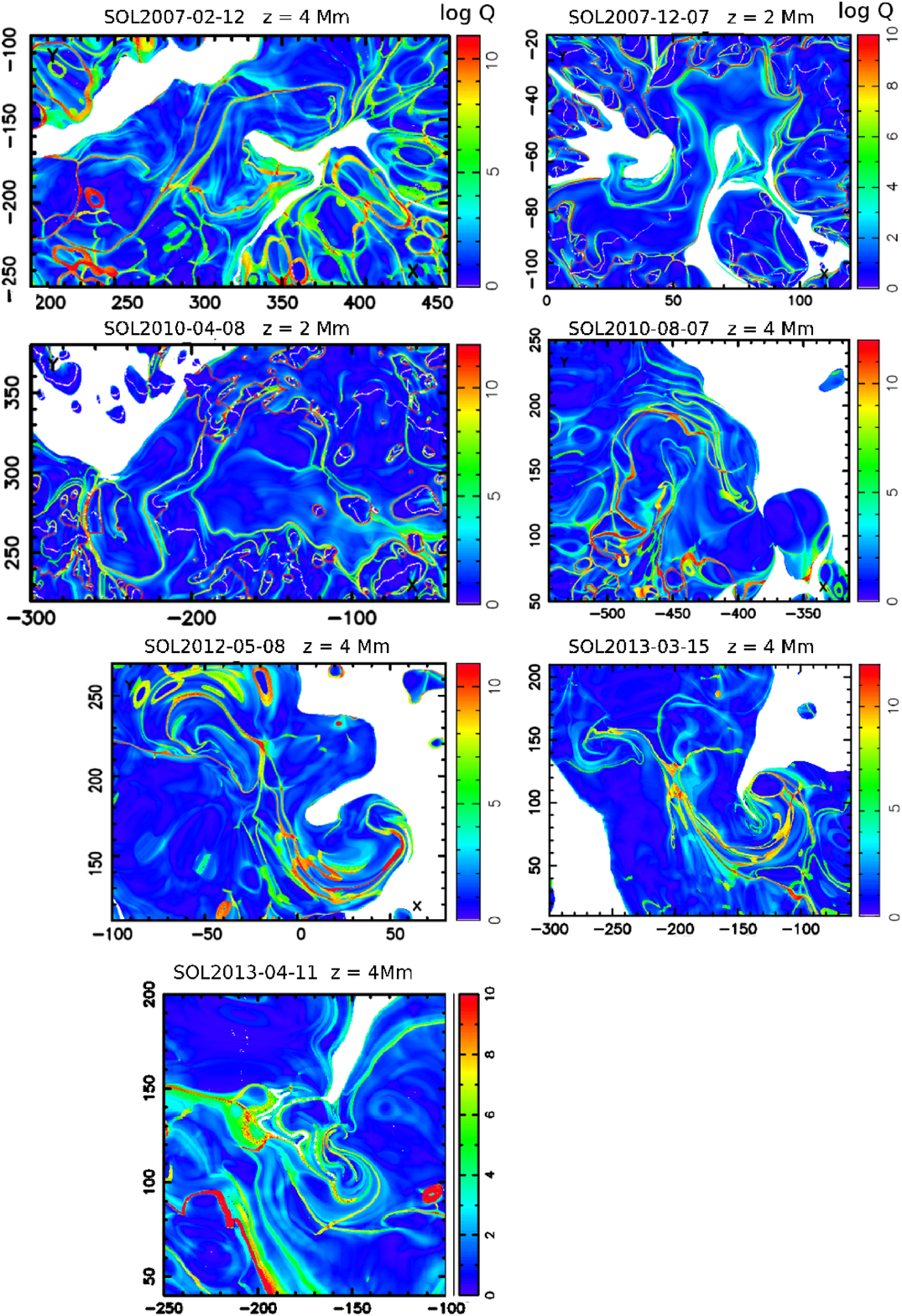}
\caption{QSL maps for all regions studied here. The QSL maps are in logarithmic scale in $Q$. The axes are given in model unites, in which one unit in $x$ or $y$ is 1\,Mm. This corresponds roughly 1.5$^{\prime\prime}$ for the purpose of comparing with the observations figures. Notice the J and S-shaped QSLs associated with the flux ropes. 
\label{QSLs}} 
\end{figure}

In most cases the maps at $z=2$\,Mm were still too complex with many small-scale features, although in the case of SOL2007-12-07 and SOL2010-04-08 these were the best matching maps. Therefore, we usually chose the maps at $z=4$\,Mm as the best match (the reference boundary for the computation of $Q$ always being at $z=2$\,Mm). These best fitting maps are shown in Fig.\,\ref{QSLs} in a logarithmic scale. Note that we reach values of $10^{10-12}$ in $Q$ in these maps, which makes us confident that we capture the important QSLs that bind the flux rope and quasi-separate it from the surrounding field, as shown in S12. In this figure the highest values of $Q$ are in reddish shades, the background $Q$ is blue, and areas where one end of the field line was located outside the computational domain (so $Q$ could not be computed) are white. In some cases, the QSLs are somewhat S-shaped (SOL2010-08-07) and most are 2J-shaped (SOL2007-02-12, SOL2010-04-08, SOL2012-05-08, SOL2013-03-15). 

\begin{figure}[h!]
\epsscale{0.8}
\plotone{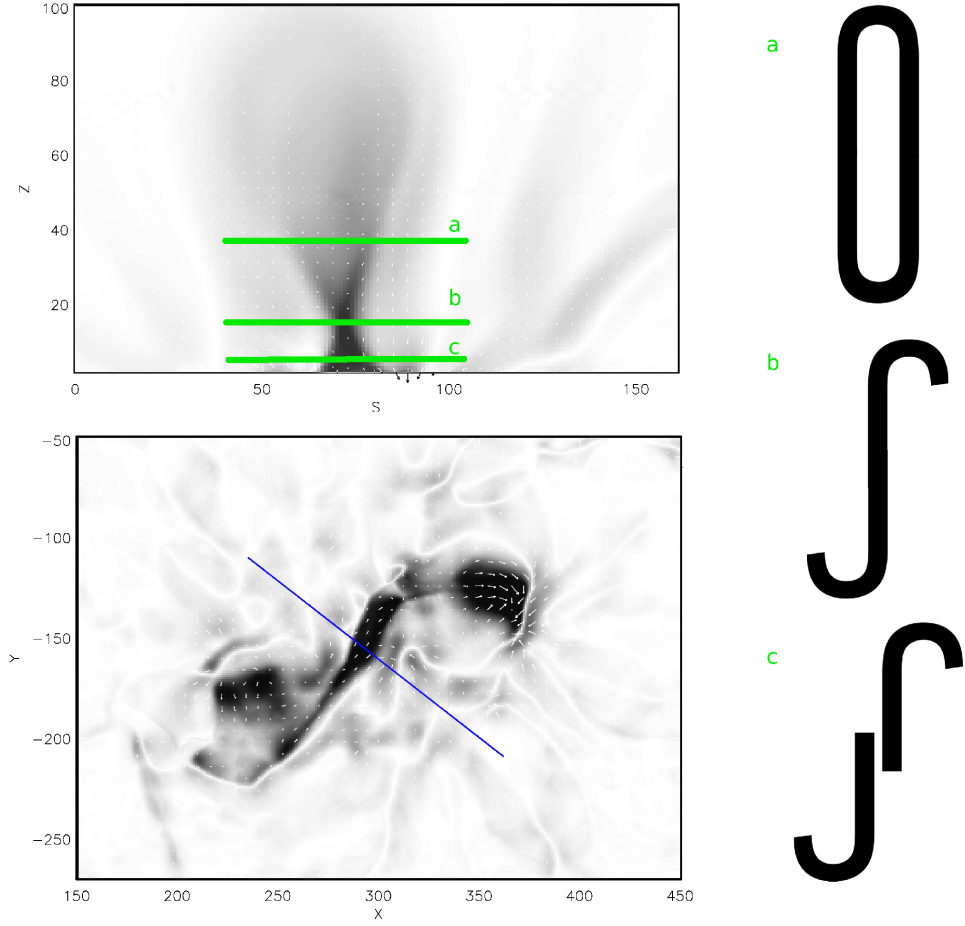}
\caption{A horizontal map showing the current in the flux rope (lower panel) and a cross-section through the flux rope at the location of the blue line in the lower panel (upper panel). The pinching under the flux rope at the HFT is visible. The synthetic shape of the QSLs is shown in the right depending on where the horizontal cut is taken with respect to the HFT (shown schematically on the vertical cross-section through the flux rope).  
\label{scheme}} 
\end{figure}

The characteristic 3D shape of the QSLs associated with a twisted flux rope is presented in in Fig.\,\ref{scheme} \citep[see also][]{Titov02, Savcheva12b,Pariat12}. Fig.\,\ref{scheme} presents a synthetic description of the QSL model of the SOL2007-02-12 sigmoid. We show the electric current density distribution in a horizontal cut at level 4\,Mm and at the location of the blue line -- a current cross-section through the flux rope. We use the current as proxy for the locations for QSLs since in S12, we showed there is a very close correspondence although the current is more diffuse than the QSLs. The pinching under the flux rope is apparent in the figure where the HFT is. It is evident that since this is an unstable model, the HFT is already at some height above the photosphere. In green we show the location of three horizontal cuts through the flux rope, and on the right the schematic view of the QSL shapes are given depending on where the cut is taken. From this it is evident that if the horizontal cut is taken at the HFT, the QSL will look S-shaped, and if the cut is below it, it will look 2J-shaped, i.e with the straight parts of the Js parallel to each other and the PIL, and the hooks facing each other at the far ends. All the maps that do not show S-shaped QSLs show 2J-shaped QSLs corresponding to the cut under the HFT. 

The correspondence between current layers and QSLs is further demonstrated in Fig.\,\ref{JQ_overlays}, where we show the horizontal distribution at the height of the best-fit QSL maps with the QSLs overlaid. The QSLs are selected to be above the same threshold as in Fig.\,\ref{QSL_overlays_1} but also to be in areas where the magnetic field is more than 15-30\,G for the different regions. This filtering allows us to remove a large part of the QSLs that are created from small scale polarities. Compared with Fig.\,\ref{QSLs}, we highlight only the QSLs that are going though intense magnetic polarities, or where energy and currents are likely to be stronger. It is clear that indeed most QSLs overlay the edges of the current concentrations. This is to be expected since, as it can be seen from Fig.\,\ref{scheme} top panel, the domain under the HFT is filled with intense current bound by the legs of the HFT. As shown in S12 and in paper II, most flux ropes show a hollow core current distribution, meaning that the current is concentrated on the edge of the flux rope. Then, in the flux rope, these ridges in the current are overlaid by the flux rope-binding QSL that separates it from the overlaying field. However, it is important to note that QSL theory does not expect an exact correspondence between the maxima in the current concentrations and QSLs since the exact relationship between the current and QSLs is complicated, not analytically determined, and dependent on the exact motions in the photosphere \citep{Aulanier06b, Wilmot09}.   

As theoretically predicted, our magnetic field models show the most intense currents accumulate where the connectivity gradients are strongest. We showed previously in S12 that the flux rope has a hollow core current distribution with current concentrated at the edge, as can also be seen from Fig.\,\ref{scheme}. This current concentration coincides with the QSL that wraps around the flux rope, which is the case for all regions that we study. In \cite{Janvier13}, this correspondence was demonstrated in the 3D MHD simulation and in \cite{Janvier15} for data-constrained QSLs and observed current ribbons.

\section{Flare Ribbons and CME Topologies} \label{sec:QSLCME}

Several studies have shown the correspondence between photospheric QSLs and ribbons in case by case examples (see Sect. \ref{sec:Intro}). Here, for the first time, we show good-quality matches between QSLs and flare ribbons in seven different regions containing flux ropes that exhibit classical two-ribbon flares.

\begin{figure}[h!]
\epsscale{1.0}
\plotone{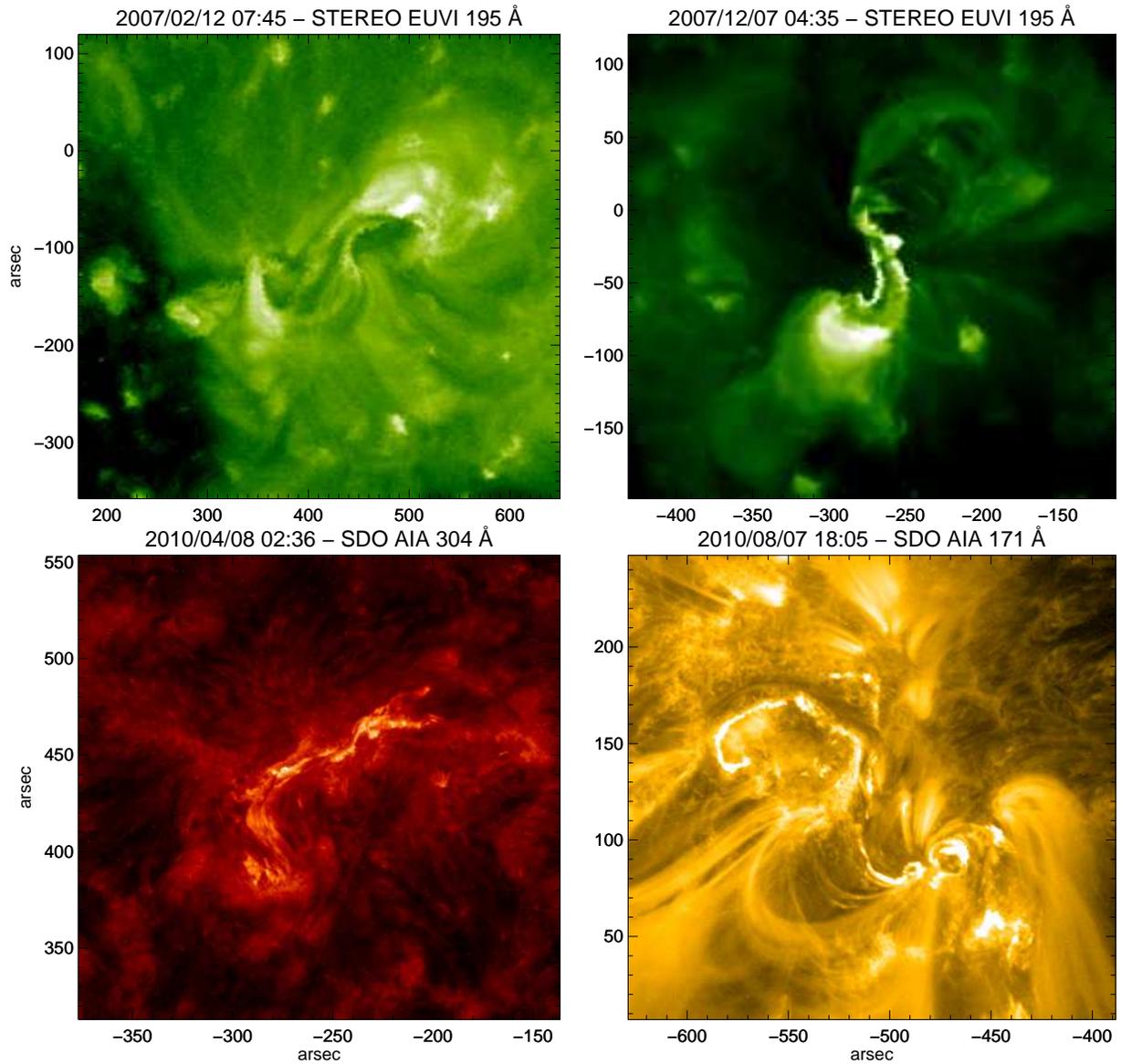}
\caption{STEREO (when AIA is not available) and AIA images for four of the regions showing the flare ribbons at $t_{ribbon}$, i.e. the first moment when they appear (see text about how we determine this moment). These images are used to match to the QLSs in Fig.\,\ref{QSL_overlays_1}.  
\label{Ribbons_AIA_1}} 
\end{figure}

\begin{figure}[h!]
\epsscale{1.0}
\plotone{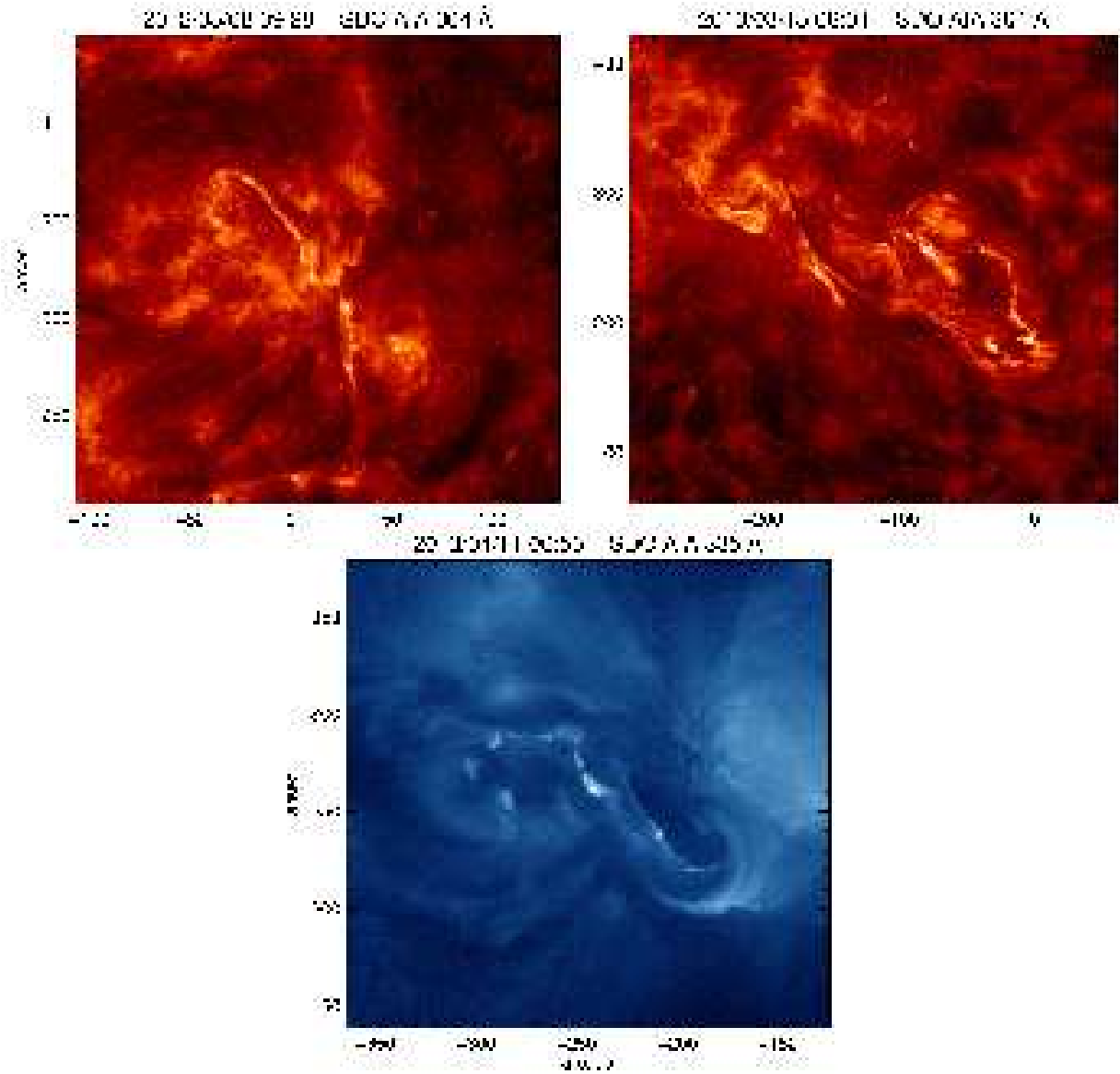}
\caption{AIA images for all regions showing the flare ribbons at $t_{ribbon}$. These images are used to match to the QLSs in Fig.\,\ref{QSL_overlays_2}.  
\label{Ribbons_AIA_2}} 
\end{figure}

\begin{figure}[h!]
\epsscale{0.8}
\plotone{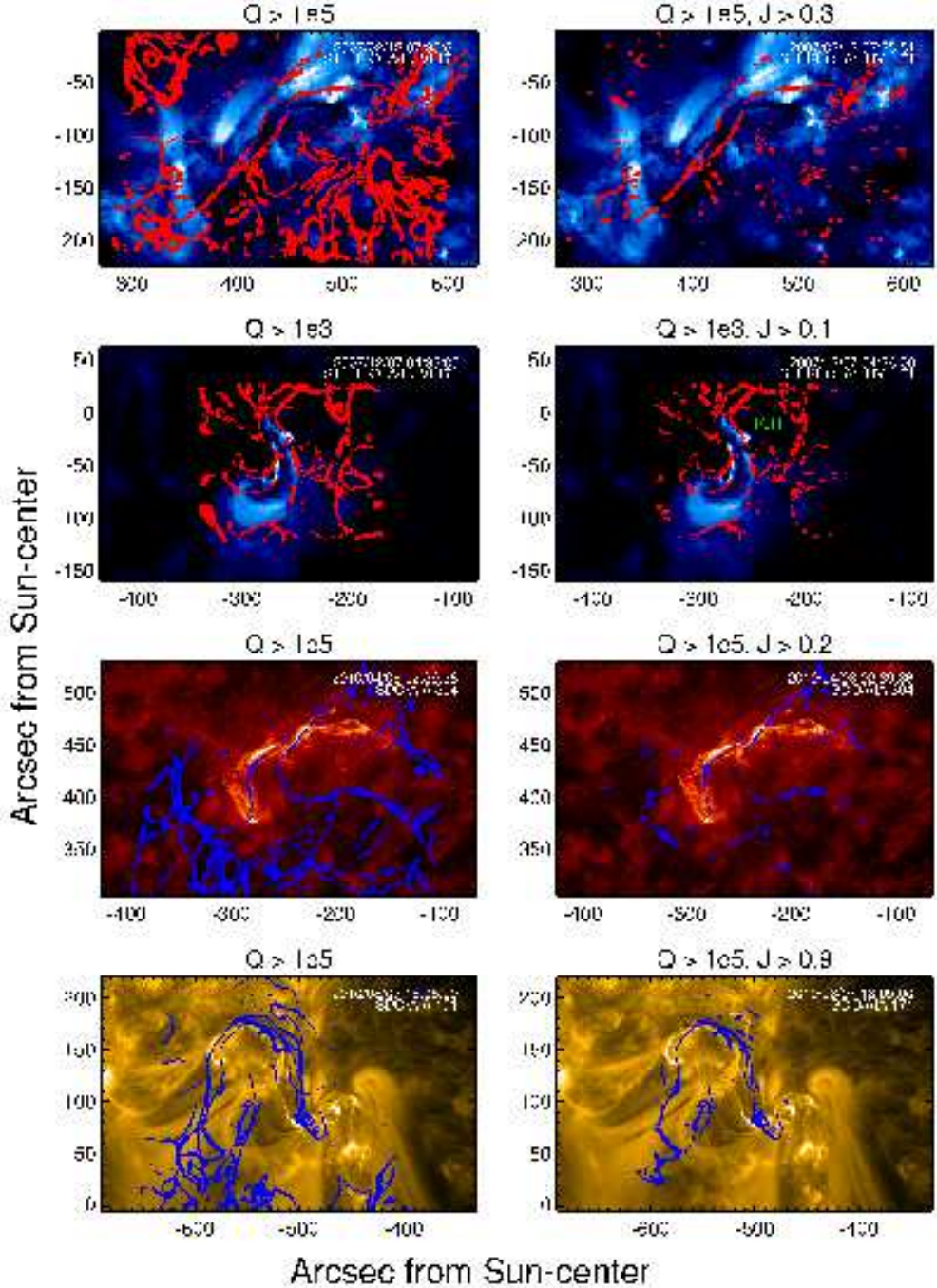}
\caption{Overlays between the flare ribbons (underlying image) shortly after the beginning of the flare and the QSL maps (red or blue curves) for the studied regions. All regions with $Q_{thresh}>10^6$ are shown in the left column. In the right column we have plotted only the QSLs which are above the same values of $Q$ but we have also required that the current density is above some value (usually $j_{thresh}>0.2$).
\label{QSL_overlays_1}} 
\end{figure}

\begin{figure}[h!]
\epsscale{0.9}
\plotone{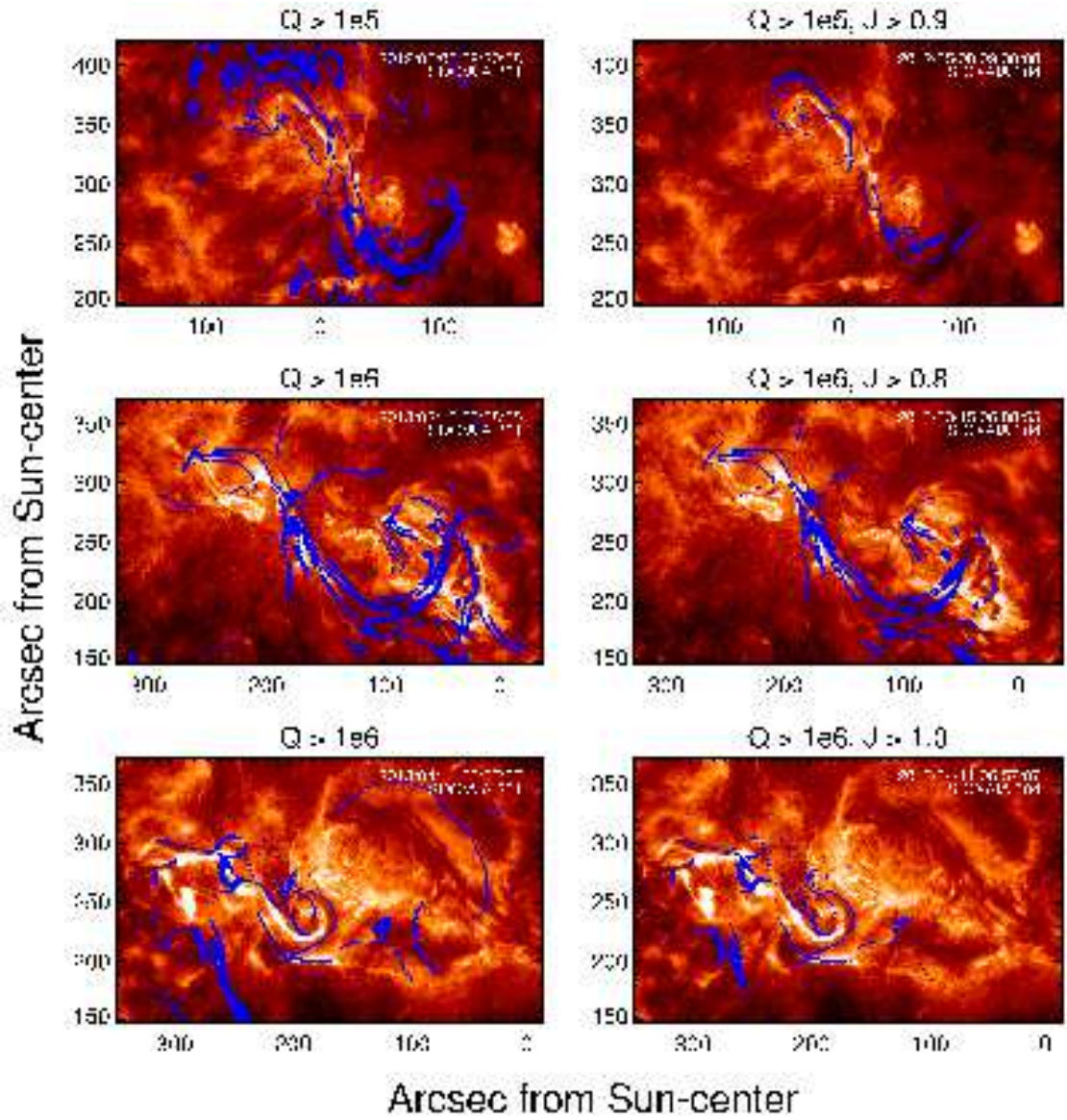}
\caption{Same as Fig.\,\ref{QSL_overlays_1} for the rest of the regions.
\label{QSL_overlays_2}} 
\end{figure}

In Fig.\,\ref{QSL_overlays_1} and \ref{QSL_overlays_2}, we show the matches between the QSLs (from Fig.\ref{QSLs}) and the flare ribbons (from Fig.\ref{Ribbons_AIA_1} and \ref{Ribbons_AIA_2}). In the left columns, the overlays between the images and all QSLs with $Q$ values above $10^{5-6}$ are displayed (the $Q$ threshold is marked above each panel). In the right columns, we overlay the flare images with QSLs with $Q$ values only in areas where the current is above a certain threshold (the $j$ threshold is also marked above each panel). 

\begin{figure}[h!]
\epsscale{0.8}
\plotone{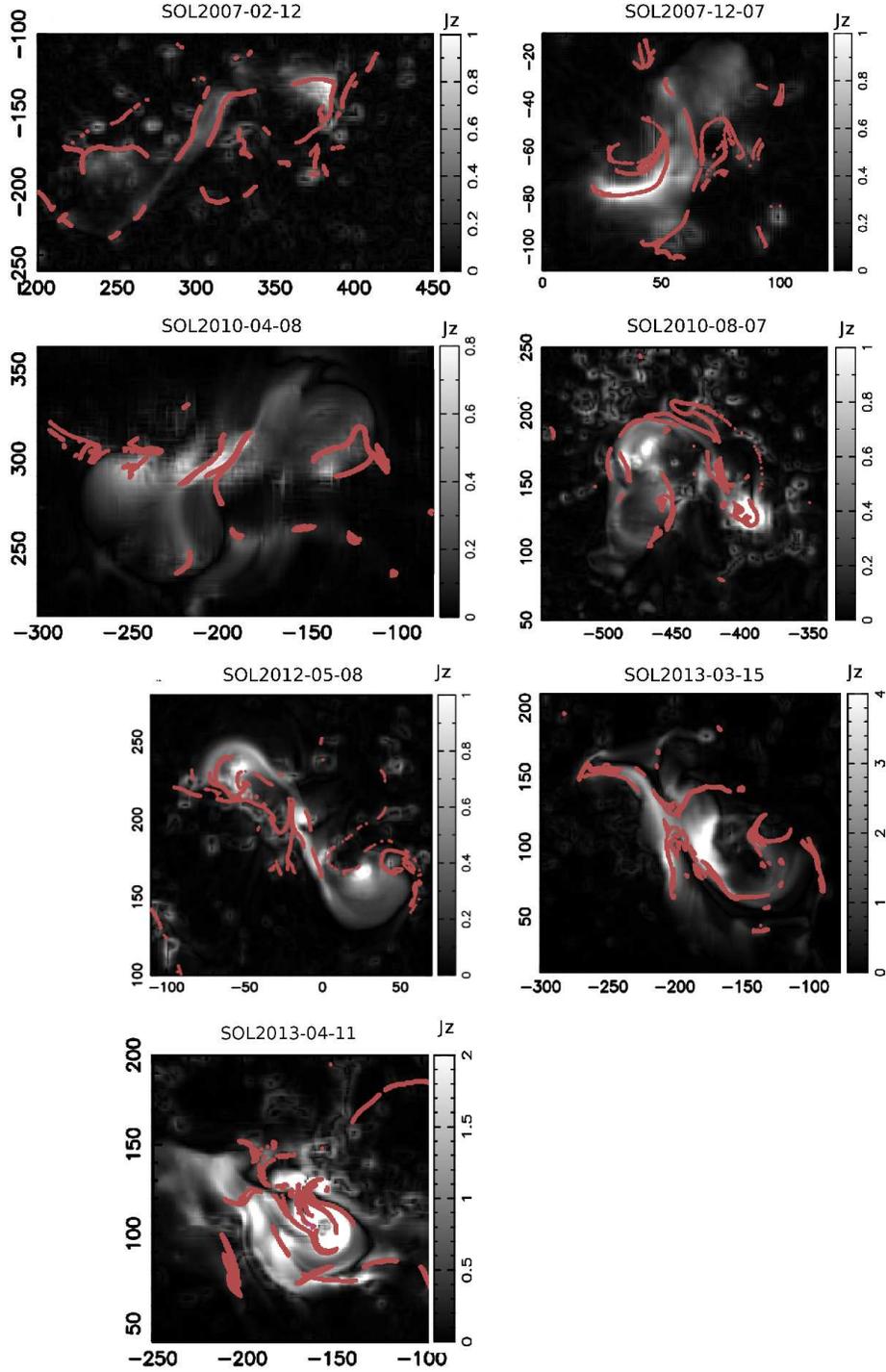}
\caption{Overlays between the vertical current density distribution in greyscale at the height of the best-fit QSL maps overlaid with QSLs with $Q>10^6$ (red lines). The QSL are only plotted in areas with LoS magnetic field larger than 15-30\,G. 
\label{JQ_overlays}} 
\end{figure}

In the following, we will not discuss whether a given QSL is associated with a given flare ribbon. In the real observed regions, while a QSL may be present, no current may accumulate at these QSLs and thus no flaring activity will be observed. We will thus mainly discuss whether an observed ribbon is associated with a QSL or not. The use of the threshold on the electric currents allows us to display mainly the QSLs along which currents have accumulated in sheets (Fig.\ref{JQ_overlays}). As discussed in Section\,4, these QSLs are most likely to host magnetic reconnection and hence show flare ribbons. By using this selection criterion, we highlight the QSLs that our models predict to be the preferential sites for the appearance of flare ribbons. A posteriori, we do observe that the overlays with the current criterion (right panels of Fig.\ref{QSL_overlays_1} and \ref{QSL_overlays_2}) give a better correspondence between the ribbons and the QSLs than with the magnetic strength criterion as in Fig.\ref{JQ_overlays}. In the left panels of Fig.\ref{QSL_overlays_1} and \ref{QSL_overlays_2}, numerous high-$Q$ regions are observed to be associated with no EUV emission. In the real observed regions, while a QSL was actually present, the dynamics did not lead to any current accumulation at these QSLs and thus no flaring activity was observed. The better correspondence obtained with the right panels is a further demonstration of the overall predictive capabilities of the marginally unstable modeling method.  

The height at which the best-matching QSL maps are taken vary from $z=2-4$\, model cells or 2100\,km to 4200\,km. The times at which the overlays are made are those shown in Fig.\,\ref{Ribbons_AIA_1} and \ref{Ribbons_AIA_2}, i.e. the first instance, $t_{ribbons}$, at which the whole ribbons appear. The iteration at which the QSL maps are taken are between 30,000 and 60,000 for the unstable models that are closest to the best-fit model. We will show in Paper II that as the relaxation proceeds, an unstable model shows an elevating HFT with legs of that move apart in time (relaxation iteration) and map the motion of the expanding flare ribbons. 

Based on the criteria for the quality of the match between the ribbons and the QSLs, we now discuss each active region emphasizing the parts of the ribbons that are well captured as well as the parts where the QSLs fail to follow the ribbons, along with some possible reasons why. Particularly, we look for a match in the shape of the ribbons and QSLs in the straight (parallel to the PIL) part as well as the hooks. The separation of the QSLs/ribbons depends on the height of the of the reconnection site and the geometry of the legs of the HFT. These control the separation of the footprints of the HFT at chromospheric heights, where the ribbons are produced. The observations do not give us any information about the height of the reconnection site at the onset of the flare. Hence, we simply choose an iteration from the magnetofrictional evolution of the flux rope that has the HFT at the appropriate height to match the initial ribbon separation from the observations. As we show in paper II, the subsequent ribbon separation in time is matched by the rising HFT in the models. The evolution of the HFT with iteration number depends mostly (if not solely) on the amount of axial flux. The poloidal flux, according to the standard 3D model, rather, controls the tightness of the hooks but not the separation of the ribbons in the part parallel to the PIL. For a future study, we have planned to explore in detail the influence of different initial parameters on the shape of the QSLs. In this study, however, we restrict ourselves only to showing the closest marginally stable or unstable model to the best-fit pre-eruption model, which contains the predictive power of this method.  

\begin{itemize}
\item {\bf SOL2007-02-12T07:30:00L075C297} (Fig.\,\ref{QSL_overlays_1}, first row): This is the same region we studied in S12, the time evolution of which was explored in \cite{Savcheva09, Savcheva12c}. For this weak B-class flare, the QSLs do not match well the flare ribbons, in particular in the east part of the sigmoid. This is probably  because the flux rope path that we insert is too long, since it could not be constrained by the STEREO images very well. The endpoints of the dark EUV filament were not well discernible. The weak strength of the flare also corresponds to a limited intensity of the flare brightennings. We again emphasize that in all cases we only use the pre-eruptive observational information to constrain the models, but in this case the match could be improved if one uses the round transient CHs, with center location around $(x,y)=$(470,-90) and (380, -160), to constrain the anchor points of the flux rope better. The presence of weaker magnetic fields compared to other studied regions also reduced the constraint on the magnetic field during the magnetofrictional relaxation. Fields lines are more free to connect in various places of the domain than in a region with stronger field. This region thus represents a case which is below the lower limit, in terms of flare class and magnetic field strength, to what one is able to model with a good correspondence.

\item {\bf SOL2007-12-07T04:20:00L085C296} (Fig.\,\ref{QSL_overlays_1}, second row): This is another relatively weak flare (B2) in a relatively quiet region. These ribbons are very well matched by the computed QSLs, which almost exactly overlay both the parallel parts of the J-ribbons around $x=-270$ and the hooks around (-280, 0) for the Northern ribbon and $x=-300$ for the Southern ribbon. The $y$ coordinate goes from -120 to 25. In addition, a QSL binds the region marked ``TCH'' for transient coronal hole in the corresponding image in Fig.\,\ref{QSL_overlays_1}, which is where the flux rope northern endpoint is situated and the dimming is observed as reduced emission. In paper II, we will explore whether the dimming regions also match the QSLs outlining the footprints of the flux ropes. This analysis has large implications to extending the standard flare model to 3D in realistic coronal magnetic fields.

\item {\bf SOL2010-04-08T02:35:00L110C176} (Fig.\,\ref{QSL_overlays_1}, third row): This is a B4 class flare occurring in a region with more intense magnetic fields. The central straight part of the ribbons, the bars of the two J's around (-250, 440), are very well matched in this region. Although a nice eastern hook is visible in the QSL map around (-330, 375), the observed ribbons display a much narrower hook, i.e. the bar of the J and the straight part of the hook of the J are very close to each other around (-300, 400). This can clearly be seen in the movie provided in the online version. The hooks become more prominent at later times. Later, the eastern hook is relatively wide (but not as much as the one in the QSLs) and the western one is also visible. This region was extensively modeled by \cite{Su11}, and the model is very well constrained. However, it is possible that the surrounding field limits the extension of the flare ribbons, while this behavior is not seen in the QSL. The dynamics of the QSLs in this region will be discussed extensively in Paper II.

\item {\bf SOL2010-08-07T17:55:00L100C002} (Fig.\,\ref{QSL_overlays_1}, fourth row): This M class flare is also overall very well matched by the model. The ribbons in this region resemble more an S-shape rather than two Js. Both the north-east hook around (-580, 160), and the more parallel part around (-500, 130) of the ribbons are well represented with only slight deviations. The swirly ribbon coinciding with the sunspot at (-480, 70) in the south-west end of the main ribbon is not captured by the QSL map since field lines that originate in this area leave the computational domain and $Q$ cannot be calculated for them; these areas appear white in the corresponding QSL map in Fig.\,\ref{QSLs}. In the left panel that contains all QSLs, there are two QSLs that pass near the north-east hook, but after the threshold on the current is imposed (in the right panel) only one remains. These two QSLs spread later on and overlay the two spreading flare ribbons, as we will discuss in Paper II.

\item {\bf SOL2012-05-08T09:21:00L108C232} (Fig.\,\ref{QSL_overlays_2}, first row): For this C class flare, the north-east hook around (-40, 375) as well as the straight part, centered at (10, 320), of the 2J ribbons are well matched by QSLs. However, there is basically no south-west ribbon hook where the QSL clearly forms a hook around (90, 240) at this early time. Later, if one refers to the movie in the online version one can see a hook forming at the location of the western hook in the QSLs. This region's NLFFF model was constrained by an almost straight, slightly S-shaped filament observed in AIA 304\AA\ that clearly coincided with the S-shaped loops seen in XRT before the flare. However, as the eruption developed another branch of this filament lying to the east of the southern end of the first filament became apparent (with its Easternmost end at round (-60, 200)). A ribbon associated with the lift-off of this other filament appears in the images around $x=-60$ to $x=50$ and $y=220$, which is not associated with any QSLs. We performed alternative models, with different flux rope paths, but were unsuccessful at reproducing this part of the ribbon. The details in this modeling will be discussed in Paper II.
          
\item {\bf SOL2013-03-15T05:46:00L116C077} (Fig.\,\ref{QSL_overlays_2}, second row): This M class flare displays a complicated pattern of ribbons with a lot of complexity in the basic shape of the ribbons and possibly more than two ribbons, but nonetheless a 2J-shape is clearly discernible. These 2J-shaped ribbons are approximately well matched by the QSLs. The best match is found in the parallel part of the ribbons centered at (-180, 240). The hooks around (-230, 320) and (-60, 200) are more poorly matched due to their higher complexity, which is missed by the QSLs. The photospheric magnetic field distribution of this region, and SOL2013-04-11, is significantly more complex. These two regions also produced the strongest flares in our sample (M-class flares). These facts might be the reason for the complex shape of the ribbons. It is also possible that the overlying envelope of the sigmoid is not potential, but there is significant non-potentiality away from the flux rope, which can cause some of the mismatch.  

\item {\bf SOL2013-04-11T06:55:00L110C071} (Fig.\,\ref{QSL_overlays_2}, third row): The situation with this region is very similar to SOL2013-03-15, i.e. strong flare, complex magnetic field distribution, and possible non-potentiality away from the main flux rope. However, in this case both the straight part (centered at (-240, 270)) and the hooks (around (-300, 290) and (-190, 220)) are very well matched, including the western hook which wraps around the sunspot at (190, 235). The extent of the ribbons is well matched as well as the little curvy protrusion in the eastern ribbon at (-250, 290).
\end{itemize}

Overall, the QSLs match the flare ribbons very well in six cases out of seven. This is a success for our approach since we have to deal with large complexity in the data-constrained QSL maps and the observed flare ribbons. In addition, these models are only constrained by the observed pre-flare loops and no information about the eruptive features such as ribbons, transient coronal holes, or flare loops, are used, although this would likely improve the match in some cases as we discussed above. 

In most cases, we find that the straight part of all ribbons is very well matched by the QSLs. Both hooks match well in four of the regions, and all regions have one hook matching. The hooks are generally harder to reproduce because this requires the most precise determination of the locations of the footprints of the filament that we use to guide our flux rope insertion. In addition, the shape of the hooks is dependent on the twist of the flux rope, as discussed above, and our models are less sensitive to the initial amount of poloidal flux in the flux rope, so the twist is less accurately determined.  
 
In the case where we do not have a very good match between the ribbons and QSLs, the model is not well constrained by the data before the eruption. The flare in this region is also the weakest and appears in highly dispersed magnetic polarities. 

Since we have used only pre-eruptive information about the coronal structure and surface magnetic field, the above, largely successful, comparisons between ribbons and QSLs show the predictive power of our method. We can infer the eruptive topology and related features with a good success based only on the pre-eruptive parameters. This link will become even stronger in paper II, where we show that this way we can also match the observed evolution of the ribbons with time and the initial configuration and the strong-to-weak shear transition in the flare loops. 

\section{Discussion and Conclusions} \label{sec:Conclusions}

In this paper, we showed the first comprehensive study of a selection of seven flaring sigmoidal active regions (SOL2007-02-12, SOL2007-12-07, SOL2010-04-08, SOL2010-08-07, SOL2012-05-08, SOL2013-03-15, and SOL2013-04-11) that display flare ribbons, transient coronal holes, and flare loops during solar flares. We select the regions purely based on the observations, i.e having 2 J-shaped flare ribbons that appear in sigmoidal active regions and have a very good coverage of the pre-flare and flare features. Here we put the ribbons in a topological context, utilizing quasi-separatrix layer (QSL) maps obtained based on data-constrained magnetic field models. In order to compute the topology in each of the seven erupting sigmoidal regions, we use the flux rope insertion method to produce marginally stable and unstable magnetic field models to derive the 3D magnetic field structure at the flare onset. The use of unstable models in a magnitofrictional approach is justified by the fact that when used as initial conditions in an MHD simulation these unstable models evolve in a very similar manner as in the subsequent iterations of the magnetofrictional relaxation process \citep[see][for details]{Kliem13}. In this sense we use only observational information about the pre-eruptive magnetic field configuration and to predict the location, shape and extent of flare- and CME-related features. The QSL computation is performed in the same manner as in \cite{Savcheva12b} and \cite{Pariat12}. 

We show an very good match between QSL maps taken at low heights in the corona and chromosphere and the locations, extent, and shape of flare ribbons at the beginning of the flares in six out of seven cases that cover a wide range of flare intensities. Generally, the straight part of the J-shaped ribbons that are parallel to the PIL are better matched in all cases. To match the hooks, one requires  very detailed knowledge of the location of the footpoints of the flux rope and the amount of twist contained in it. In this sense, our models are successful at reproducing both hooks in four of the cases. SOl2007-02-12 does not display a very good match in the eastern hook due to the poorly constrained path of the pre-eruptive flux rope. This study presents the largest sample of correspondence of active region flare ribbons and topological features derived from data-constrained magnetic field models. 

Being confident that our models match the observed flare features gives us the ability to analyze the flaring configurations. We can go further by using the ``requirement'' that the flare ribbons match the QSLs and apply it as a constraint on our models in order to understand the 3D reconnection process. For example, one can determine the height of the HFT at different locations along the flux rope and the location of the flare. One can perform torus instability diagnostics, as we did in \cite[][, S12]{Savcheva12b}, knowing the shape, size, and center of the flux rope from the QSL cross-sections. One can look at the connectivities of field lines and their change during the eruptions. We reserve most of this analysis for Paper II, where we will perform extensive dynamics analysis on several consecutive iterations of the relaxation process and show the evolution of the QSLs and field line connectivities with iteration number (time). The analysis given in this paper and in Paper II has potentially very strong implications for the study of 3D reconnection in realistic coronal magnetic fields, which depart from  the usual symmetry that many reconnection simulations and laboratory experiment employ.   

Previous studies have used LFFF and potential models to match observed flare features to separatrices and QSLs (see Sect.\,\ref{sec:Intro}). Recently \cite{Liu14} and \cite{Zhao14} used stable NLFFF models of complex regions with flux ropes. However, they did not interpret the connection between the flare ribbons and low-lying QSLs in the same way as we do here. While such NLFFF methods are useful for studying magnetically complex regions that do not show a clear flux rope evidence, they are largely limited in their applicability (flux rope systems) as discussed in \cite{Schrijver06}, and they are intrinsically tied to the pre-flare configuration, since the surface magnetograms do not usually change significantly in response to the flare. On the other hand, by using unstable flux rope configuration in a magnetofrictional method we have the ability to follow the rise of the flux rope as the flare (reconnection) progresses, although we caution that this is not a self-consistent approach. We simply use it as a simpler model that maps the 3D magnetic field structure of the observed phenomena surprisingly well. To be fully consistent, one needs to use a data-driven MHD simulation or a time dependent magnetofrictional approach, which is planned for a future study to confirm this result.

Most of our knowledge about reconnection comes from effectively 2D reconnection experiments, and only recently efforts to understand 3D magnetic reconnection have been pursued \citep{Priest11, Pontin11, Shepherd12, Janvier14}, revealing a much wider range of dynamics. Very little is known about the actual properties of the solar plasma during the reconnection process on the Sun. Many authors have studied  the global and local reconnection rates and timescales of solar reconnection during solar flares based on analysis of the ribbon motions in the chromosphere \citep{Fletcher01,Xie09,Qiu10,Miklenic07,Qiu02,Saba06,Isobe05,Jing05}. Generally, they use measurements of the magnetic flux at the photosphere, the velocity of the ribbons parallel to the PIL, and the 2D approximation of the standard flare model to determine the reconnected flux and the electric field and Poynting flux at the reconnection site. In a future paper, we plan to test the validity of this approximation. However, these studies are purely based on the ribbon dynamics and are not able to capture the dynamics of the 3D coronal magnetic field, which is actually involved in the reconnection process. For example, to calculate the energy release rate from reconnection, one needs to combine observations with knowledge of the location and size (i.e. length, the width is generally under the resolution) of the reconnection current sheet, which can only be provided by data-constrained magnetic field models and extrapolations or by data-driven MHD simulations with the appropriate NLFFF initial conditions (and is still only an approximation). \cite{Aulanier00, Aulanier12} and \cite{Schrijver11} utilize idealized MHD simulations to interpret the appearance of the flare ribbons. Although such studies show a qualitative similarity of the QSLs and flare ribbons and are valuable for determining the basic topology of the region, no actual estimates of the reconnection parameters can be obtained, which could be improved by the use of data-constrained and data-driven magnetic field modeling. 

So far, dynamics studies of the reconnection process have been either purely observational (reconnection rates, reconnected flux, etc.) or based on idealized MHD simulations. A quasi-dynamic (as in the current approach), dynamic magnetofrictional (from sequence of magnetograms as in \cite{Gibb14}), or a data-driven MHD simulation with a NLFFF initial condition \citep[as in][]{Kliem13} will prove to be more valuable in the study of the evolving ribbons as the flare progresses and ultimately in the study of the process and properties of observed 3D reconnection on the Sun.    

\acknowledgments

{\it Hinode} is a Japanese mission developed, launched, and operated by ISAS/JAXA in partnership with NAOJ, NASA, and STFC (UK). Additional operational support is provided by ESA, NSC (Norway). This work was supported by NASA contract NNM07AB07C to SAO. The QSL computations have been performed on the multi-processors TRU64 computer of the LESIA. Savcheva is supported by the NASA LWS Jack Eddy postdoctoral fellowship. Su is supported by NSFC \# 11333009 and Youth Fund of Jiang Su \# BK20141043.  We would like to thank the AIA team for supplying the data for this study.


\bibliographystyle{apj}  
\bibliography{Antonia_sig_topo}       
\IfFileExists{\jobname.bbl}{}  
{ 
\typeout{} 
\typeout{****************************************************} 
\typeout{****************************************************} 
\typeout{** Please run "bibtex \jobname" to obtain}  
\typeout{**the bibliography and then re-run LaTeX}  
\typeout{** twice to fix the references!} 
\typeout{****************************************************} 
\typeout{****************************************************} 
\typeout{} 
 } 
 
\clearpage

\begin{table}\scriptsize
\begin{center}
\caption{Summary of the timing of observed flare features for all seven regions.\label{tbl-1}}
\begin{tabular}{ccccccccc}
\tableline\tableline
ID & Event & GOES class & $t_{GOES}$ & Ribbons & $T_{ribbons}$ & Transient CHs & Flare loops \\ 
& & & & AIA\,304\AA\,/ & & AIA\,193\AA\,/ & AIA\,171\AA\,/ \\ 
& & & & STEREO\,195\AA\, & & STEREO\,195\AA\, & STEREO\,195\AA\,\\
\tableline
SOL2007-02-12 & 12 Feb 2007 & $<$B & 07:40 & 07:35-11:55 & 07:35 & 07:45-11:55 & 09:05-10:45 \\
SOL2007-12-07 & 07 Dec 2007 & B2 & 04:20 & 04:35-05:15 & 04:36 & 04:25-05:15 & 04:55-06:45 \\
SOL2010-04-08 & 08 Apr 2010 & B4 & 02:35 & 02:20-06:00 & 02:39 & 03:02-06:00 & 02:49-06:00 \\
SOL2010-08-07 & 07 Aug 2010 & M1 & 17:55 & 17:52-20:27 & 18:05 & 18:12-19:02 & 18:36-20:31 \\
SOL2012-05-08 & 08 May 2012 & C1.7 & 09:21 & 09:22-13:00 & 09:30 & 09:41-10:16 & 09:56-12:58 \\
SOL2013-03-15 & 15 Mar 2013 & M1.1 & 05:46 & 05:57-09:00 & 06:08 & 06:11-gap & gap-09:00 \\
SOL2013-04-11 & 11 Apr 2013 & M6.5 & 06:55 & 06:27-08:00 & 06:57 & 07:00-08:00 & 07:24-08:00 \\
\tableline
\end{tabular}
\end{center}
\end{table} 
 
\clearpage

\begin{table}\tiny
\begin{center}
\caption{Best-fit model parameters for all modeled regions.\label{tbl-2}}
\begin{tabular}{ccccccccc}
\tableline\tableline
ID & Event date & $t_{mag}$ & $\Phi_{axi}$ & $F_{pol}$ & $F_{pol,tot}$ & $E_{pot}$ & $E_{free}$ & $h_{rel}$ \\ 
& & & $[$10$^{20}$ Mx] & [10$^{10}$ Mx cm$^{-1}$] & [10$^{20}$ Mx] & [10$^{31}$ erg] & [10$^{31}$ erg] & [10$^{42}$ Mx$^2$]\\ 
\tableline
SOL2007-02-12 & 12 Feb 2007 & 06:41 & 5 & 5 & 3.5 & 3.64 & 1.47 & 1.08 \\
SOL2007-12-07 & 07 Dec 2007 & 03:21 & 7 & 5 & 1.1 & 1.28 & 0.59 & 0.34 \\
SOL2010-04-08 & 08 Apr 2010 & 02:00 & 6 & 1 & 4.1 & 5.29 & 1.82 & 1.85 \\
SOL2010-08-07 & 07 Aug 2010 & 17:00 & 15 & -0.5 & -3.6 & 29.7 & 8.22 & -6.88 \\
SOL2012-05-08 & 08 May 2012 & 05:38 & 7 & -1 & -4.0 & 10.5 & 2.83 & -2.53 \\
SOL2013-03-15 & 15 Mar 2013 & 05:00 & 5 & -0.5 & -2.4 & 53.8 & 7.34 & -6.61 \\
SOL2013-04-11 & 11 Apr 2013 & 04:50 & 7 & -0.5 & -2.5 & 41.2 & 9.26 & -4.78 \\
\tableline
\end{tabular}
\end{center}
\end{table}

\clearpage

\clearpage

\clearpage

\clearpage

\clearpage

\clearpage

\clearpage

\end{document}